# Incorporating molecular data in fungal systematics: a guide for aspiring researchers

Kevin D. Hyde[1,2], Dhanushka Udayanga[1,2], Dimuthu S. Manamgoda[1,2], Leho Tedersoo[3,4], Ellen Larsson[5], Kessy Abarenkov[3], Yann J.K. Bertrand[5], Bengt Oxelman[5], Martin Hartmann[6,7], Håvard Kauserud[8], Martin Ryberg[9], Erik Kristiansson[10], R. Henrik Nilsson[5,*]

[1] Institute of Excellence in Fungal Research

[2] School of Science, Mae Fah Luang University, Chiang Rai, 57100, Thailand

[3] Institute of Ecology and Earth Sciences, University of Tartu, Tartu, Estonia

[4] Natural History Museum, University of Tartu, Tartu, Estonia

[5] Department of Biological and Environmental Sciences, University of Gothenburg, Box 461, 405 30 Göteborg, Sweden

[6] Forest Soils and Biogeochemistry, Swiss Federal Research Institute WSL, Birmensdorf, Switzerland

[7] Molecular Ecology, Agroscope Reckenholz-Tänikon Research Station ART, Zurich, Switzerland

[8] Department of Biology, University of Oslo, PO Box 1066 Blindern, N-0316 Oslo, Norway

[9] Department of Organismal Biology, Uppsala University, Norbyvägen 18D, 75236 Uppsala, Sweden

[10] Department of Mathematical Statistics, Chalmers University of Technology, 412 96 Göteborg, Sweden

[*] Corresponding author: R. Henrik Nilsson – email – henrik.nilsson@bioenv.gu.se

## Abstract

The last twenty years have witnessed molecular data emerge as a primary research instrument in most branches of mycology. Fungal systematics, taxonomy, and ecology have all seen tremendous progress and have undergone rapid, far-reaching changes as disciplines in the wake of continual improvement in DNA sequencing technology. A taxonomic study that draws from molecular data involves a long series of steps, ranging from taxon sampling through the various laboratory procedures and data analysis to the publication process. All steps are important and influence the results and the way they are perceived by the scientific community. The present paper provides a reflective overview of all major steps in such a project with the purpose to assist research students about to begin their first study using DNA-based methods. We also take the opportunity to discuss the role of taxonomy in biology and the life sciences in general in the light of molecular data. While the best way to learn molecular methods is to work side by side with someone experienced, we hope that the present paper will serve to lower the learning threshold for the reader.

## Introduction

Morphology has been the basis of nearly all taxonomic studies of fungi. Most species were previously introduced because their morphology differed from that of other taxa, although in many plant pathogenic genera the host was given a major consideration (Rossman and Palm-Hernández 2008; Hyde et al. 2010; Cai et al. 2011). Genera were introduced because they were deemed sufficiently distinct from other groups of species, and the differences typically amounted to one to several characters. Groups of genera were combined into families and families into orders and classes; the underlying reasoning for grouping lineages was based on single to several characters deemed to be of particular discriminatory value. However, the whole system hinged to no small degree on what characters were chosen as arbiters of inclusiveness. These characters often varied among mycologists and resulted in disagreement and constant taxonomic rearrangements in many groups of fungi (Hibbett 2007; Shenoy et al. 2007; Yang 2011).

The limitations of morphology-based taxonomy were recognized early on, and numerous mycologists tried to incorporate other sources of data into the classification process, such as information on biochemistry, enzyme production, metabolite profiles, physiological factors, growth rate, pathogenicity, and mating tests (Guarro et al. 1999, Taylor et al. 2000; Abang et al. 2009). Some of these attempts proved successful. For instance, the economically important plant pathogen *Colletotrichum kahawae* was distinguished from *C. gloeosporioides* based on physiological and biochemical characters (Correll et al. 2000; Hyde et al. 2009). Similarly, relative growth rates and the production of secondary metabolites on defined media under controlled conditions are valuable in studies of complex genera such as *Aspergillus, Penicillium,* and *Colletotrichum* (Frisvad et al. 2007; Samson and Varga 2007; Cai et al. 2009). In many other situations, however, these data sources proved inconclusive or included a signal that varied over time or with environmental and experimental conditions. In addition, the detection methods often did not meet the requirement as an effective tool in terms of time and resource consumption (Horn et al. 1996; Frisvad et al. 2008).

Molecular data was first used in taxonomic studies of fungi in the 1970's (cf. DeBertoldi et al. 1973), which marked the beginning of a new era in fungal research. The last twenty years have seen an explosion in the use of molecular data in systematics and taxonomy, to the extent where many journals will no longer accept papers for publication unless the taxonomic decisions are backed by molecular data. DNA sequences are now used on a routine basis in fungal taxonomy at all levels. Molecular data in taxonomy and systematics are not devoid of problems, however, and there are many concerns that are not always given the attention they deserve (Taylor et al. 2000; Groenwald et al. 2011; Hibbett et al. 2011). We regularly meet research students who are unsure about one or more steps in their nascent molecular projects. Much has been written about each step in the molecular pipeline, and there are several good publications that should be consulted regardless of the questions arising. What is missing, perhaps, is a wrapper for these papers: a freely available, easy-to-read yet not overly long document that summarizes all major steps and provides references for additional reading for each of them. This is an attempt at such an overview. We have divided it into six sections that span the width of a typical molecular study of fungi: 1) taxon sampling, 2) laboratory procedures, 3) sequence quality control, 4) data analysis, 5) the publication process, and 6) other observations. The target audience is research students ("the users") about to undertake a (Sanger sequencing-based) molecular mycological study with a systematic or taxonomic focus.

## 1. Sampling and compiling a dataset

The dataset determines the results. If there is anything that the user should spend valued time completing, it is to compile a rich, meaningful dataset of specimens/sequences (Zwickl and Hillis 2002). Right from the onset, it is important to consider what hypotheses will be tested with the resulting phylogeny. If the hypothesis is that one taxon is separate from one or several other taxa, then all those taxa should be sampled along with appropriate outgroups; consultation of taxonomic expertise (if available) is advisable. It is paramount to consider previous studies of both the taxa under scrutiny and closely related taxa to achieve effective sampling. It is usually a mistake to think that one knows about the relevant literature already, and the users are advised to familiarize themselves with how to search the literature efficiently (cf. Conn et al. 2003 and Miller et al. 2009). In particular, it often pays off to establish a set of core publications and then look for other papers that cite these core papers (through, e.g., “Cited by” in Google Scholar). If there is an opportunity to sequence more than one specimen per taxon of interest, this is highly preferable and particularly important when it comes to poorly defined taxa and/or at low taxonomic levels. In most cases it will not be enough to use whatever specimens the local herbarium or culture collection has to offer; rather the user should consider the resources available at the national and international levels. Ordering specimens or cultures may prove expensive, and the user may want to consider inviting researchers with easy access to such specimens as co-authors of the study. The invited co-authors may even oversee the local sequencing of those specimens, to the point where more specimens does not have to mean more costs for the researcher. When considering specimens for sequencing, it should be kept in mind that it may be difficult to obtain long, high-quality DNA sequences from older specimens (cf. Larsson and Jacobsson 2004). However, there are other factors than age, such as how the specimen was dried and how it has been stored, that also influence the quality of the DNA. There also seem to be systematic differences across taxa in how well their DNA is preserved over time; as an example, species adapted to tolerate desiccation are often found to have better-preserved DNA than have short-lived mushrooms (personal observation). There are different methods that increase the chances of getting satisfactory sequences even from single cells or otherwise problematic material (see Möhlenhoff et al. 2001, Maetka et al. 2008, and Bärlocher et al. 2010).

*Sampling through herbaria, culture collections, and the literature.* Mining world herbaria for species and specimens of any given genus is not as straightforward as one might think. Many herbaria (even in the Western countries) are not digitized, and no centralized resource exists where all digitized herbaria can be queried jointly. GBIF (http://www.gbif.org/) nevertheless represents a first step in such a direction, and the user is advised to start there. Larger herbaria not covered by GBIF may have their own databases searchable through web interfaces but otherwise are best queried through an email to the curator. Type specimens have a special standing in systematics (Hyde and Zhang 2008; Ko Ko et al. 2011; MacNeill et al. 2012), and the inclusion of type specimens in a study lends extra weight and credibility to the results and to unambiguous naming of generated sequences. Herbaria may be unwilling to loan type specimens (particularly for sequencing or any other activity that involves destroying a part of the specimen, i.e., destructive sampling) and may offer to sequence them locally instead – or may in fact already have done so. Specimens collected by particularly professional taxonomists or that are covered in reference works should similarly be prioritized. For cultures, we recommend the user to start at the CBS culture collection (http://www.cbs.knaw.nl/collection/AboutCollections.aspx) and the StrainInfo web portal (http://www.straininfo.net/). Not all relevant specimens are however deposited in public collections; many taxonomists keep personal herbaria or cultures. Scanning the literature for relevant papers to locate those specimens is rewarding. It may be particularly worthwhile to try to include specimens reported from outlying localities, in unusual ecological configurations, or together with previously unreported interacting taxa with respect to those specimens already included. Such exotic specimens are likely to increase both the genetic depth and the discovery potential of the study. Collections from distant locations and substrates other than that of the type material may represent different biological species. Although they cannot always be readily distinguished based on morphology (i.e., they are cryptic species) including these taxa will allow for a better understanding of the targeted species.

Incorrectly labelled specimens will be found even in the most renowned herbaria and culture collections. The fact that a leading expert in the field collected a specimen is only a partial safeguard against incorrect annotations, because processes other than taxonomic competence – such as unintentional label switching and culture contamination - contribute to misannotation. However most herbaria and culture collections welcome suggestions and accurate annotations (with reasonable verifications) for specimens that are not identified correctly or are ambiguous. It is a good idea to double check all specimens retrieved and to seek to verify the taxonomic affiliation of the specimens in the sequence analysis steps (sections 3 and 4). Keeping an image of the sequenced specimen is valuable.

*Sampling through electronic resources.* As more and more researchers in mycology and the scientific community sequence fungi and environmental samples as part of their work, the sequence databases accumulate significant fungal diversity. Even if the users know for certain that they are the only active researchers working on some given taxon, they can no longer rest assured that the databases do not contain sequences relevant to the interpretation of that taxon. On the contrary, chances are high that they do. As an example, Ryberg et al. (2008) and Bonito et al. (2010) both used insufficiently identified fungal ITS sequences in the public sequence databases to make significant new taxonomic and geo/ecological discoveries for their target lineages. Therefore, it is recommended that sequences similar to the newly generated sequences should be retrieved in order to establish phylogenetic relationships and verify the accuracy of the sequence data through comparison.

This process is simple and amounts to regular sequence similarity searches using BLAST (Altschul et al. 1997) in the International Nucleotide Sequence Databases (INSD: GenBank, EMBL, and DDBJ; Karsch-Mizrachi et al. 2012) or *emerencia* (http://www.emerencia.org) – details on how to do these searches and what to keep in mind when interpreting BLAST results are given in Kang et al. (2010) and Nilsson et al. (2011, 2012). In short, the user must not be tempted to include too many sequences from the BLAST searches. We advise the user to focus on sequences that are very similar to, and that cover the full length of, the query sequences. If the user follows this approach for all of their newly generated sequences, then they should not have to worry about picking up too distant sequences that would cause problems in the alignment and analysis steps. Whether the sequences obtained through the data mining step are annotated with Latin names or not – after all, about 50% of the 300,000 public fungal ITS sequences are not identified to the species level (cf. Abarenkov et al. 2010) – does not really matter in our opinion. They represent samples of extant fungal biodiversity, and they carry information that may prove essential to disentangle the genus in question. The user should keep an eye on the geo/ecological/other metadata reported for the new entries to see if they expand on what was known before.

A word of warning is needed on the reliability of public DNA sequences. As with herbaria and culture collections, many of these sequences carry incorrect species names (Bidartondo et al. 2008) or may be the subject of technical problems or anomalies. Sequences stemming from cloning-based studies of environmental samples are, in our experience, more likely to contain read errors, or to be associated with other quality issues, than sequences obtained through direct sequencing of cultures and fruiting bodies. The process of establishing basic quality and reliability of fungal DNA sequences, including cloned ones, is discussed further in section 3.

*Field sampling and collecting.* Many fungi cannot be kept in culture and so can never be purchased from culture collections. Most herbaria are biased towards fungal groups studied at the local university or groups that are noteworthy in other regards, such as economically important plant pathogens. The public sequence databases are perhaps less skewed in that respect, but instead they manifest a striking geographical bias towards Europe and North America (Ryberg et al. 2009). There are, in other words, limits to the fungal diversity one can obtain from resources already available, such that collecting fungi in the field often proves necessary. Indeed, collecting and recording metadata are cornerstones of mycology, and it is essential, amidst all digital resources and emerging sequencing technologies, that we keep on recording and characterizing the mycobiota around us in this way (Korf 2005). Such voucher specimens and cultures form the basis of validation and re-determination for present and future research efforts regardless of the approach adopted. In addition, some morphological characters can only be observed or quantified properly in fresh specimens, alluding further to the importance of field sampling. (Descriptions of such ephemeral characters should ideally be noted on the collection.)

Fruiting bodies should be dried through airflow of ≤40 degrees centigrade or through silica gel for tiny specimens. The dried material should be stored in an airtight zip-lock (mini-grip) plastic bag to prevent re-wetting and access by insect pests. (Insufficiently dried material may be degraded by bacteria and moulds, leaving the DNA fragmented and contaminated by secondary colonizers.) It is recommended to place a piece of the (fresh) fruiting body in a DNA preservation buffer such a CTAB (hexadecyl-trimethyl-ammonium bromide; PubChem ID 5974) solution; ethanol and particularly formalin-based solutions perform poorly when it comes to DNA preservation and subsequent extraction (cf. Muñoz-Cadavid et al. 2010). In addition to fruiting bodies, it is recommended to store any vegetative or asexual propagules such as mycelial mats, mycorrhizae, rhizomorphs, and sclerotia that the user intends to employ for molecular identification. CTAB buffer works fine for these too. If DNA extraction is planned in the immediate future, samples of collections can be placed into DNA extraction buffer already in the field. For fresh samples that have no soil particles, a modified Gitchier buffer (0.8M Tris-HCl, 0.2 M $(NH_4)_2SO_4$, 0.2% w/v Tween-20) can be used for cell lysis and rapid DNA extraction (Rademaker et al. 1998).

With respect to plant-associated microfungi, plant specimens should be used for fungal isolation as soon as possible after they are collected from the field, otherwise they can be used for direct DNA extraction when still fresh. In the case of leaf-inhabiting fungi, the plant material should be dried and compressed using standard methods, without any toxicogenic preservatives or treatments. For plant-pathogenic fungi and culturable microfungi, the practice of single-spore isolation is desirable (Choi et al. 1999; Chomnunit et al. 2011). Indeed, monosporic/haploid cultures/material are recommended whenever possible in molecular taxonomic studies and several other contexts, such as genome sequencing. Heterokaryotic mycelia or other structures may manifest heterozygosity (i.e., co-occurring, divergent allelic variants) for the targeted marker(s), which would add an unwanted layer of complexity in many situations. Methods for single spore isolation, desirable media for initial culturing, and preservation of cultures are equally important factors to consider with respect to improvements of the quality in molecular experiments (see Voyron et al. 2009 and Abd-Elsalam et al. 2010).

Hibbett et al. (2011) made a puzzling observation on the limits of known fungal diversity: when fungi from herbaria are sequenced and compared to fungi recovered from environmental samples (e.g., soil), these groups form two more or less disjoint entities. By sequencing one of them, we would still not know much about the other. Porter et al. (2008) similarly portrayed different scenarios for the fungal community at a site in Ontario depending on whether aboveground fruiting bodies or belowground soil samples were sequenced. The fact that a non-trivial number of fungi do not seem to form (tangible) fruiting bodies has been known for a long time, but it is essential that field sampling protocols consider this. One idea could be to increase the proportion of somatic (non-sexual) fungal structures sampled during field trips (cf. Healy et al. in press). Right now that proportion may be close to zero, based on the last few organized field trips we have attended. Such collecting should be seen as a long-term project unlikely to yield results immediately, but we suggest it would be a good idea if the users, when out collecting for some project, would try to collect, voucher, and sequence at least one fungal structure they would normally have ignored.

**2. Selection of gene/marker, primers, and laboratory protocols**

The protocols pertaining to the laboratory work come with numerous decisions, many of which will fundamentally affect the end results. The best way to learn laboratory work is through someone experienced, such that the users do not have to make all those decisions on their own. Many pitfalls and mistakes can be avoided in this way. A sound step towards making informed choices also involves looking into the literature: what genes/markers, primers, and laboratory protocols were used by researchers who studied the same or closely related taxa (with roughly similar research questions in mind)? That said, many scientific studies come heavily underspecified in the Materials & Methods section, and the user should look to it for inspiration rather than for full recipes. After selecting target marker(s) and appropriate primers, the user should be prepared to spend time in the molecular laboratory for one to several weeks. The laboratory processes involved in a typical molecular phylogenetics study is DNA extraction, a PCR reaction to amplify the gene(s)/marker(s), examination and purification of the PCR products, the sequencing reactions, and the screening of the resulting fragments.

*Choice of gene/marker.* It is primarily the research questions that dictate what genes or markers to target. For resolution at and below the generic level (including species descriptions), the nuclear ribosomal ITS region is a strong first candidate. The ITS region is typically variable enough to distinguish among species, and its multicopy nature in genomes makes it easy to amplify even from older herbarium specimens or in other situations of low concentrations of DNA. The ITS region is the formal fungal barcode and the most sequenced fungal marker (Begerow et al. 2010; Schoch et al. 2012). For some groups of fungi, other genes or markers give better resolution at the species level (see Kauserud et al. 2007 and Gazis et al. 2011). As single molecular markers, GPDH and Apn2/MAT work well for the *Colletotrichum gloeosporioides* complex (Silva et al. 2012, Weir et al. 2012); GPDH for the genera *Bipolaris* and *Curvularia* (Berbee et al. 1999, Manamgoda et al. 2012); MS204 and FG1093 for *Ophiognomonia* (Walker et al. 2012); and tef-1α for *Diaporthe* (Castlebury et al. 2007; Santos et al. 2010, Udayanga et al. 2012). For research questions above the genus level, the user should normally turn to other genes than the ITS region. The nuclear ribosomal large subunit (nLSU) has been a mainstay in fungal phylogenetic inference for more than twenty years – such that a large selection of reference nLSU sequences are available – and largely shares the ease of amplification with the ITS region. It is challenged, and often surpassed, in information content by genes such as β-tubulin (Thon and Royse 1999), tef-1α (O'Donnell et al. 2001), MCM7 (Raja et al. 2011), RPB1 (Hirt et al. 1999), and RPB2 (Liu et al. 1999). As a general observation, single-copy genes are typically more difficult to amplify from small amounts of material and from moderate-quality DNA than are multi-copy genes/markers such as nLSU and ITS (Robert et al. 2011; Schoch et al. 2012). Genes known to occur as multiple copies (e.g., β-tubulin) in certain fungal genomes may produce misleading systematic conclusions based on single-gene analysis (Hubka and Kolarik 2012).

For larger phylogenetic pursuits it is common, even standard, to include more than one unlinked gene/marker in phylogenetic endeavours (James et al. 2006). Phylogenetic species recognition by genealogical concordance, typically relying on more than one gene genealogy, has become a tool of modern day systematics of species complexes of fungi and other organisms (Taylor et al. 2000; Monaghan et al. 2009; Leavitt et al. 2011). Indeed, many journals have come to expect that two or more genes be used in a phylogenetic analysis, and it may be a good idea to use, e.g., a ribosomal gene such as nLSU and a non-ribosomal gene in molecular studies. It should be kept in mind that all genes cannot be expected to work equally well in all fungal lineages, both in terms of amplification success and information content. For species descriptions and phylogenetic inferences of lesser scope it is typically still deemed acceptable – although perhaps not recommendable – to use a single gene, and for these purposes we advocate the ITS region as the primary marker due to its high information content, ease of amplification, role as the fungal barcode, and the large corpus of ITS sequences already available for comparison. However, if the new species falls outside any known genus, we recommend to also sequence the nLSU to provide an approximate phylogenetic position for the new species (lineage). Knowledge of the rough phylogenetic position of the new sequence, coupled with literature searches, may give clues to what genes that are likely to perform the best for species identification and subgeneric phylogenetic inference of the new lineage.

*Choice of primers.* When amplifying DNA from single specimens, one typically does not have to worry about whether or not the primer will match perfectly to the template. Even in case of imperfect match, the PCR surprisingly often still comes out successful. This makes it is a good idea to try standard fungal or universal primers first – such as ITS1F (forward) and ITS4 (reverse) (White et al. 1990; Gardes et al. 1993) in the case of the ITS region – because chances are high that they will work. It is however advisable to use at least one fungus-specific primer (ITS1F in the above) to reduce the chances of amplifying DNA of any co-occurring eukaryotes. The literature is likely to hold clues to what lineages require more specialized primers. For example, highly tailored primers are needed for the ribosomal genes of the agaricomycete genera *Cantharellus* and *Tulasnella* (Feibelman et al. 1994; Taylor and McCormick 2008). In the unlucky event that the standard primers do not work for the lineage targeted by the user – and nobody has developed specialized primers for that gene/marker and lineage combination already – the user may have to design new primers based on sequences in, e.g., INSD. Good software tools are available for this (e.g., PRIMER3 at http://frodo.wi.mit.edu/; Rozen and Skaletsky 1999 and Primer-BLAST at http://www.ncbi.nlm.nih.gov/tools/primer-blast/) but as indata they require some 100+ base-pairs of the regions immediately upstream and downstream of the target region. If these upstream and downstream regions are not available in the sequence databases, the user has to generate them themselves. Primers can also be designed manually based on a multiple sequence alignment (cf. Singh and Kumar 2001). Upon completing the *in silico* primer design step, the user can turn to software tools such as ecoPCR (www.grenoble.prabi.fr/trac/ecoPCR; Ficetola et al. 2010) to simulate the performance of the new primers on the target region under fairly realistic conditions. Chances are nevertheless high that the user does not need to design any new primers, particularly not if targeting any of the more commonly used genes and markers in mycology. A very rich primer array is available for the ribosomal genes (e.g., Gargas and DePriest 1996; Ihrmark et al. 2012; Porter and Golding 2012; Toju et al. 2012), and most research groups have detailed primer sections for both ribosomal and other genes and markers on their home pages (e.g., http://www.clarku.edu/faculty/dhibbett/protocols.html, http://www.lutzonilab.net/primers/index.shtml, and http://unite.ut.ee/primers.php). Several publications are available with suggestions for selecting primers for widely used molecular markers as well as recently available new markers for specific groups of fungi which are commonly researched by mycologists (Glass and Donaldson 1995, Carbone & Kohn 1999, Schmitt et al. 2009, Santos et al. 2010, Walker et al. 2012).

*Laboratory steps.* The laboratory process to obtain DNA sequences can roughly be divided into five steps: DNA extraction from the specimen, PCR, examination and purification of the PCR products, sequencing reactions, and fragment analysis/visualisation. *In vitro* cloning may sometimes be needed to pick out the correct PCR fragment for sequencing, thus forming an additional step. The last few years have seen an increasing trend of sending the purified PCR products to a commercial or institutional sequencing facility for sequencing, leaving the user to oversee only the three first of the above five steps. The present authors, too, employ such external sequencing services, and our conclusion is that it is cost- and time efficient and that the technical quality of the sequences produced is generally very high.

Some researchers prefer the traditional CTAB-based way of DNA extraction. In terms of quality of results it is a good and cheap choice (Schickmann et al. 2011). However, others find it more convenient to use one of the many commercially available kits for DNA extraction. Under the assumption that the starting material is relatively fresh and that the taxa under scrutiny do not contain unusually high amounts of compounds that affect the DNA extraction/PCR steps adversely, most extraction kits are likely to perform satisfactory at recovering enough DNA to support a PCR run (comparisons of extraction methods are available in Fredricks et al. 2005, Karakousis et al. 2006, and Rittenour et al. 2012). Substrates with high concentrations of humic acids, notably soil and wood, are known to be problematic in terms of extraction and amplification (Sagova-Mareckova et al. 2008). Similarly, high concentrations of polysaccharides, nucleases, and pigments can cause interference in extraction of DNA (and the subsequent PCR) from some genera of microfungi (Specht et al. 1982). For culturable fungi, a minimal amount (10-20 mg) of actively growing edges of cultures should be used. Rapid and efficient DNA extraction kits tend to work the best when minimal amounts of tissue are used; the use of low amounts of starting material serves to reduce the amount of potential extraction/PCR inhibitors, such that decreasing – rather than increasing – the amount of starting material is often the first thing one should try in light of a failed DNA extraction/PCR run (cf. Wilson 1997). When slow-growing culturable fungi (e.g., some bitunicate ascomycetes and marine fungi) are used in DNA extraction, the user should be mindful of the substantial time needed for the growth of the fungus in order to get a sufficient amount of tissue material for DNA extraction. One should also be aware of the risk of contamination in the extraction (as well as subsequent) steps, particularly when working with older herbarium specimens. A good rule is to never mix fresh and older fungal collections when extracting DNA and always to clean the working area and the picking tools, such as the stereomicroscope and the forceps, thoroughly before and in between each round of fungal material. The application of bleach is more efficient than, e.g.,

UV light and ethanol. It is important that PCR products never be allowed to enter the area where DNA extractions and PCR setup is performed. Avoidance of contamination is particularly important when working with closely related species, as is often done in taxonomic studies, since such contaminations are easily overlooked and may be tricky to identify afterwards.

For the PCR step, many commercial PCR kits are available on the market. One can expect a standardized, consistent performance from such kits; moreover they come with suggested PCR cycling programs under which most target DNA will amplify in a satisfactory way. However, tweaking the PCR protocol may increase the yield significantly, and it is well worth consulting with more experienced colleagues and/or any relevant publication. One particularly influential factor is the annealing temperature of the primer, which is often provided with the primer sequence or may otherwise be estimated (in, e.g., PRIMER3). It is also possible and often beneficial to perform a gradient PCR to determine the optimal annealing temperature. Various modifications and optimizations of the PCR process are discussed in Hills et al. (1996), Cooper and Poinar (2000), Qiu et al. (2001), and Kanagawa (2003). PCR runs are normally verified for success through running the products on an agarose gel stained with ethidium bromide, where any DNA sequences obtained will shine in UV light. Recently, alternative staining agents that are less toxic than ethidium bromide have been marketed as alternatives in safe handling (e.g., Goldview (Geneshun Biotech.) and Safe DNA Dye/ SYBR® Safe DNA Gel). These markers use UV light or other wavelengths for visualisation, and any new laboratory should carefully consider which system to use for detection of successful PCR products. It should be remembered that these methods may detect levels of DNA that, however low, may still suffice for successful sequencing, but it is usually a money and time saving effort to only proceed with PCR products that produce a single, clearly visible band (Figure 1). Positive and negative PCR controls should always be employed.

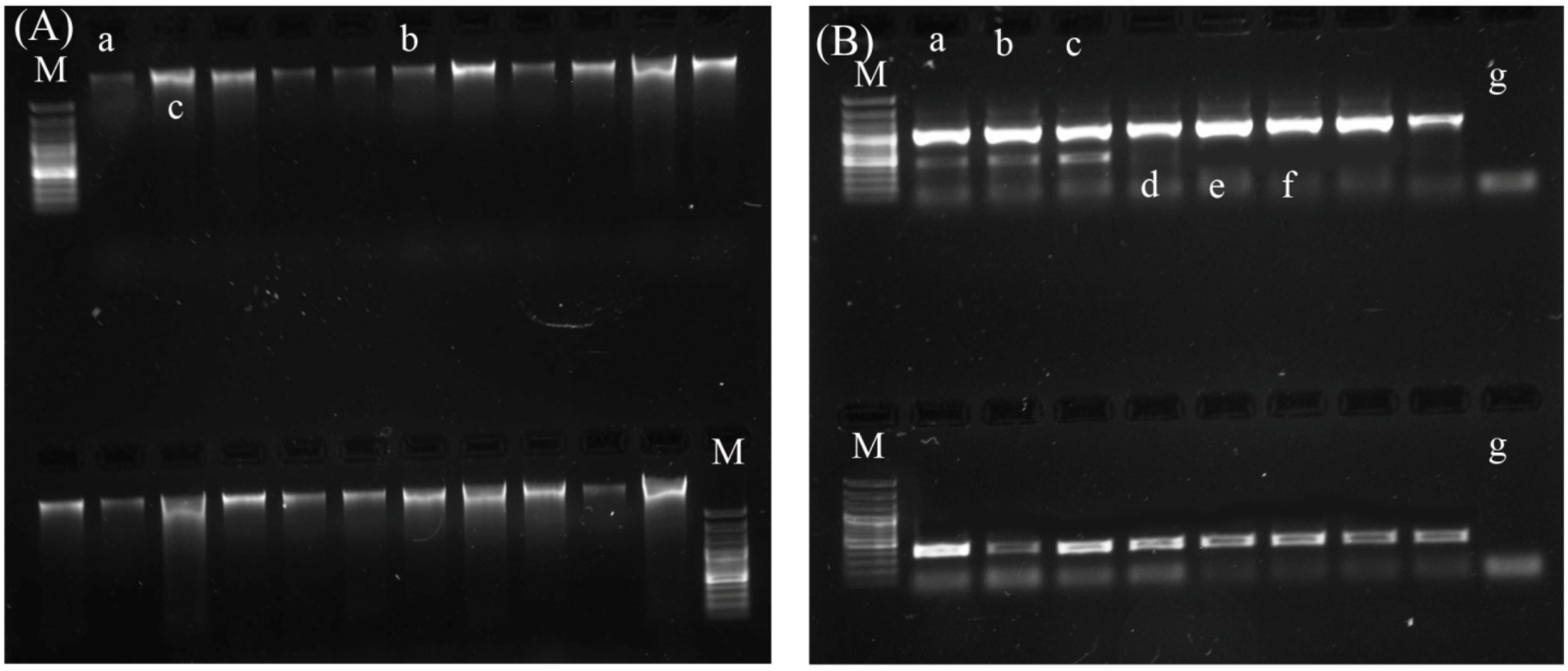

**Figure 1**. **A)** Safe-dye stained (SYBR® Safe DNA; 1% agarose gel, 80 V, 20 min) gel profiles showing a successful genomic DNA extraction of fungi. **M)** Ladders indicating the relative sizes of the DNA (100 bp marker). **a)** Low quality/minimum amount of DNA for a PCR **b)** Optimum high quality amount of DNA **c)** Excess DNA for PCR (quantification and dilution needed prior to the standard PCR reaction.)
**B)** A successful amplification of a PCR product stained with Safe dye (1% agarose gel, 80 V, 20 min) **a, b, c)** a probable case of multiple copies in amplification due to non-specific binding of primers **d, e, f)** successful amplicons with high concentrations of PCR products **g)** negative control
Courtesy: USDA-ARS Systematic Mycology and Microbiology Laboratory.

The (successful) PCR products should then be purified to remove, e.g., residual primers and unpaired nucleotides; many commercial kits are available for this purpose. These can range from single-step reactions to multiple-step, highly effective procedures. One of the simplest, cheapest, and most widely used approaches is a combination of exonuclease and Shrimp alkaline phosphatase enzymatic treatment (Hanke and Wink 1994). Before sending the purified PCR products for sequencing, they may need to be quantified for DNA content (depending on the sequencing facility). A rough quantification can be made from the strength of the band during PCR visualisation, possibly by comparing to samples of standard concentrations. Special DNA quantifiers, usually relying on (fluoro-)spectrophotometry, can be used for more exact quantification. The sequencing facilities will usually perform the sequencing reactions themselves to get optimal, tailored performance on their sequencing machine.

## 3. Sequence quality control

The responsibility to ensure that the newly generated sequences are of high authenticity and reliability lies with the user. There are many examples in the literature where compromised sequence data have lead to poor results and unjustified conclusions (cf. Nilsson et al. 2006), suggesting that the quality control step should not be taken lightly. Two points at which to exercise quality control is during sequence assembly and once the consensus sequence that has been produced from the sequence assembly is ready.

*Sequence assembly.* Many sequences in systematics and taxonomy are generated with two primers - one forward and one reverse - such that the target sequence is effectively read twice. This dual coverage brings about a mechanism for basic quality control of the read quality of the consensus sequences. The primer reads returned from the sequencing machine should be assembled in a sequence assembly program into a contig, from which the final sequence is derived (Miller and Powell 1994). Although sequence assembly is a semi-to-fully automated step in programs such as Sequencher (http://genecodes.com/), Geneious (http://www.geneious.com/), and Staden (http://staden.sourceforge.net/), the results must be viewed as tentative and need verification. The user should inspect each contig for positions (bases) of substandard appearance. During the sequencing process, the sequencing machine quantifies the light intensity of the four terminal (dyed) nucleotides at each position, and the relative intensity is represented as chromatogram curves in each primer read (Figure 2a). Occasionally the assembly software struggles to reconcile the chromatograms from the two reads, leaving the bases incorrectly determined, undecided (as IUPAC DNA ambiguity symbols such as “N” and “S”, see Cornish-Bowden 1985), or in the wrong order. The user should scan the contig and unpaired sequences along their full length for such anomalies, most of which can be identified through the odd appearance of the chromatogram curves in those positions (Figure 2b). The distal (5’ and 3’) ends of contigs are nearly always of poor quality, and the user should expect to have to trim these in all contigs. There may also be ambiguities in the chromatogram if there are different copies of the DNA region in the sequenced material. This may show as twin curves for different bases at the same site (Figure 2c). In most cases such twin curves represent heterozygous sites that should be coded using the corresponding IUPAC codes (e.g., C/T = Y). In the case of multiple heterozygous sites, it may be necessary with an extra cloning step in the lab protocol to separate the different copies. When sequencing PCR amplicons derived from dikaryotic or heterokaryotic tissue/mycelia, some sequences contigs may shift from high quality to nonsense due to the presence of an indel (insertion or deletion) in one of the alleles. In the case of only one indel present, one may obtain a usable contig sequence by sequencing the fragment in both directions.

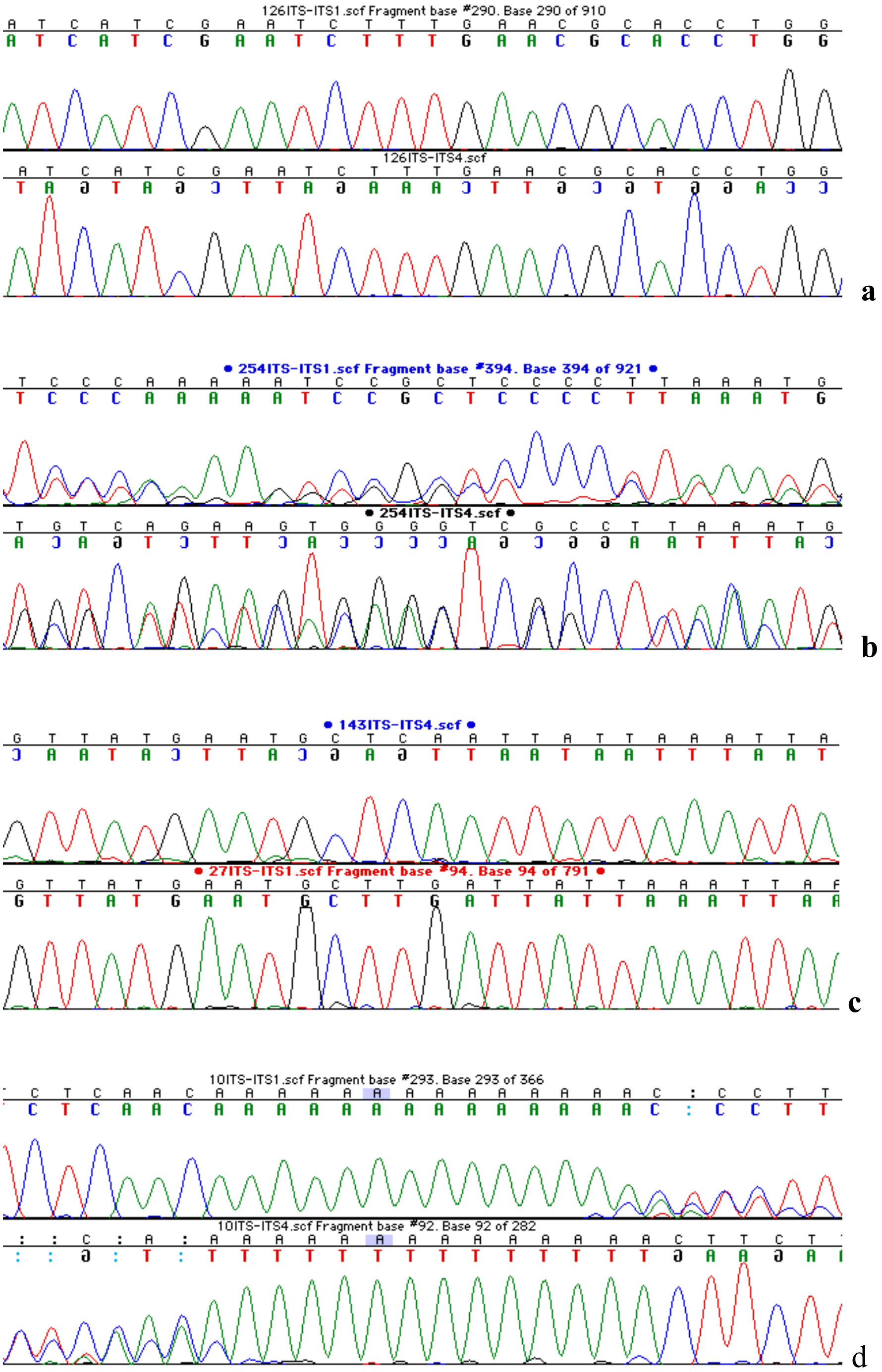

**Figure 2. a)** Clear, high-quality chromatogram curves. The software interprets curves like these with ease. **b)** Correct base-calling is hard, both for the software and for the user, when the chromatogram curves look like this. In most cases it is better to re-sequence the specimen than to try to salvage data from such chromatograms. **c)** If there is more than one (non-identical) copy of the marker amplified, the chromatogram curves tend to look like this. In the middle of the image, the uppermost primer read has produced a "TCAA" whereas the second primer produced "TTGA". The reads appear clear and unequivocal, such that poor read quality is not a likely explanation for the discrepancy. **d)** Base-calling tends to be hard in homopolymer-rich regions, and one often finds that regions after the homopolymer-rich segments are less well read. All screenshots generated from Sequencher® version 4 sequence analysis software, Gene Codes Corporation, Ann Arbor, MI USA (http://www.genecodes.com).

*A note on cloned sequences.* When performing Sanger sequencing of PCR amplicons derived directly from fruiting body tissue or mycelia, most PCR errors will normally not surface because the resulting chromatograms represent the averaged signal from numerous original templates. However, when cloning, a single PCR fragment is picked, multiplied, and sequenced. This means that any polymerase-generated errors will become visible and have to be controlled for. The only reasonable guard against such PCR generated errors is cloning and sequencing replicate fragments from the same PCR reaction. A unique mutation appearing in only one of the sequences most likely represents a PCR-generated error and should be omitted from the resulting consensus sequence. If some mutations approach a 50/50 ratio in the replicate sequences, they most likely represent allelic variants and should be analyzed separately in the further phylogenetic analyses. Although more expensive, the implementation of high-fidelity polymerase enzymes with high accuracy is especially important when sequencing cloned fragments.

*Quality control of DNA sequences.* To exercise quality control through the sequence assembly step, while very important, is only a part of the quality management process. There are many kinds of sequence errors and pitfalls that cannot be addressed during sequence assembly. Nilsson et al. (2012) listed a set of guidelines on how to establish basic authenticity and reliability for newly generated (or, for that matter, downloaded) fungal ITS sequences. Five relatively common sequence problems were addressed: whether the sequence represents the intended gene/marker, whether the sequence is given in the correct orientation, whether the sequence is chimeric, whether the sequence manifests tell-tale signs of other technical anomalies, and whether the sequence represents the intended taxon. In short, the user should never assume any newly generated sequences to be of satisfactory quality; rather, the user should take measures to ensure the basic reliability of the sequences. Such measures do not have to be complex, time-consuming, or computationally expensive. Nilsson et al. (2012) computed a joint multiple sequence alignment of their entire query ITS sequence dataset and located the conserved 5.8S gene of the ITS region in all sequences of the alignment. That approach verified that all sequences were ITS sequences and that they were given in the correct orientation. Each query sequence was then subjected to a BLAST search in INSD, and by examining the graphical BLAST summary as well as the full BLAST output, the authors were able to rule out the presence of bad chimeras and sequences with severe technical problems. Finally, for all query sequences with some sort of taxonomic annotation (e.g.,

"*Penicillium* sp."), the authors examined the most significant BLAST results for clues that the name was at least approximately right; if a sequence annotated as *Penicillium* would not produce hits to other sequences annotated as *Penicillium* (accounting for synonyms and anamorph-teleomorph relationships), then something would almost certainly be wrong. It should be kept in mind that the public sequence databases contain a non-trivial number of compromised sequences, suggesting that the guidelines – or other means of quality control – should be applied also to all sequences downloaded from such resources. The guidelines suggested do not form a 100% guarantee for high-quality DNA sequences, but they are likely to result in a more robust and reliable dataset.

## 4. Alignment and phylogenetic analysis

Systematics and tree-based thinking go hand in hand, and we urge the reader to employ a phylogenetic approach even when describing a single new species. It should be stressed, though, that phylogenetic inference is something of a research field in its own right. A phylogenetic analysis should not involve a few clicks on the mouse to produce a tree; rather it is a process that involves many decisions and that should be given significant thought. Many software packages in the field of phylogenetic inference are complex and command-line driven – intimidating, perhaps, to some biologists. But instead of resorting to simpler click-to-run programs, the user should carefully read the respective documentation and query the literature for the difference between the approaches and options. It is also worth asking for expert help or even inviting pertinent researchers as co-authors to the project. That is how important it is to get the analysis part right! Suboptimal methodological choices beget less – or incorrectly - resolved phylogenies and in the end may lead to erroneous conclusions. This is in nobody's interest.

In some cases – particularly near the species level - a phylogenetic tree may not be the most suitable model for representing the relations among the taxa in the query dataset (due to, e.g., intralocus recombination, hybridization, and introgression). In these cases, a network (Kloepper and Huson 2008) may be a better analytical solution. Networks can be used to visualize data conflicts in tree reconstruction (split networks, e.g., Bryant and Moulton 2004; Huson and Bryant 2006; Huson et al. 2011) or explicitly represent phylogenetic relationships (reticulate networks, e.g., Huber et al. 2006; Kubatko 2009; Jones et al. 2012). The user should also be aware of an ongoing paradigm shift in evolutionary biology, where single-gene trees and concatenation approaches are replaced by species tree thinking (Edwards 2009). Although the distinction between gene and species trees is not new (Pamilo and Nei 1988; Doyle 1992), most species phylogenies published to date are based on single gene trees or trees based on concatenation of two or more alignments from different linkage groups. Recently, however, models and methods have been developed to account for the fact that gene trees can differ in topology and branch lengths due to population genetics processes that are best modelled with the multispecies coalescent (Rannala and Yang 2003). In such a framework, gene trees representing unlinked parts of the cellular genomes themselves represent data for the species trees (Liu and Pearl 2007). Recent advances in sequencing technologies and whole-genome sequence information has greatly improved the ease by which multiple genes can be sampled. Interestingly, species trees also enable objective species delimitations based on molecular data (O'Meara 2010; Fujita et al. 2012). Species trees can be inferred directly from sequence data using, e.g., *BEAST (Heled and Drummond 2010).

At this stage, the user should facilitate downstream analyses by giving all of the new sequences short, unique names that feature only letters, digits, and underscore. It is good practice to start the name with a unique identifier of the sequence/specimen (since some programs truncate the names of the sequences down to some pre-defined length); this may then be followed by a more descriptive name, such as a Latin binomial. A good sequence name is *SM11c_Amanita_gemmata* ; in contrast, the default names of sequences downloaded from INSD (and that feature, e.g., pipe characters and whitespace) may cause problems in many software tools relevant to alignment and phylogenetic inference.

*Multiple sequence alignment.* Upon completing the quality management of the newly generated sequences, the user hopefully has a well-founded idea of what taxa to use as outgroups. The choice of outgroup is of substantial importance and should be done as thoroughly as possible (cf. de la Torre-Bárcena et al. 2009). Ideally, and based on the literature, at least three progressively more distantly related taxa (with respect to the ingroup) should be chosen as tentative outgroups, although it is the alignment step that decides what sequences appear suitable as outgroups and what sequences appear too distant. One regularly sees scientific publications that employ too distant outgroups; this translates into alignment problems and potentially compromised inferences of phylogeny.

There are many software tools available for multiple sequence alignment, but we urge the user to go for a recent (and readily updated) one rather than relying on old programs, well-known as they may be. We recommend any of MAFFT (Katoh and Toh 2010), Muscle (Edgar 2004), and PRANK (Löytynoja and Goldman 2005) for large or otherwise non-trivial sequence datasets. Unless the number of sequences reaches several hundreds, these programs can all be run in their most advanced mode (e.g., “linsi” in the case of MAFFT) on a regular desktop computer. The product is a multiple sequence alignment file, usually in the FASTA format (Pearson and Lipman 1988). It is however important to recognize that manual inspection of alignment files is always needed and manual adjustment is often warranted (Figure 3a). This involves loading the alignment in an alignment viewer such as SeaView (Gouy et al. 2010) and trying to find and correct for instances where the multiple sequence alignment program seems to have performed suboptimally (Figure 3b,c). Exact guidelines on how to do this are hard to give (cf. Gonnet et al. 2000, Lassmann and Sonnhammer 2005), and the user is advised to sit down with someone experienced with this process to have it demonstrated. Alternatively, the user could send the alignment for improvement to someone experienced and then contrast the two versions. Occasionally, alignments feature sections that defy all attempts at reconstruction of a meaningful alignment (Figure 3d). Such sections should be kept in the alignment and may be excluded from the subsequent phylogenetic analysis. There are several software solutions that attempt to formalize this procedure (e.g., Talavera and Castresana 2007; Liu et al. 2009; Capella-Gutierrez et al. 2009). It should be noted, though, that there may be unambiguously aligned subsets of sequences (taxa) in such globally unalignable regions, and these sub-alignments may contain useful information.

a.

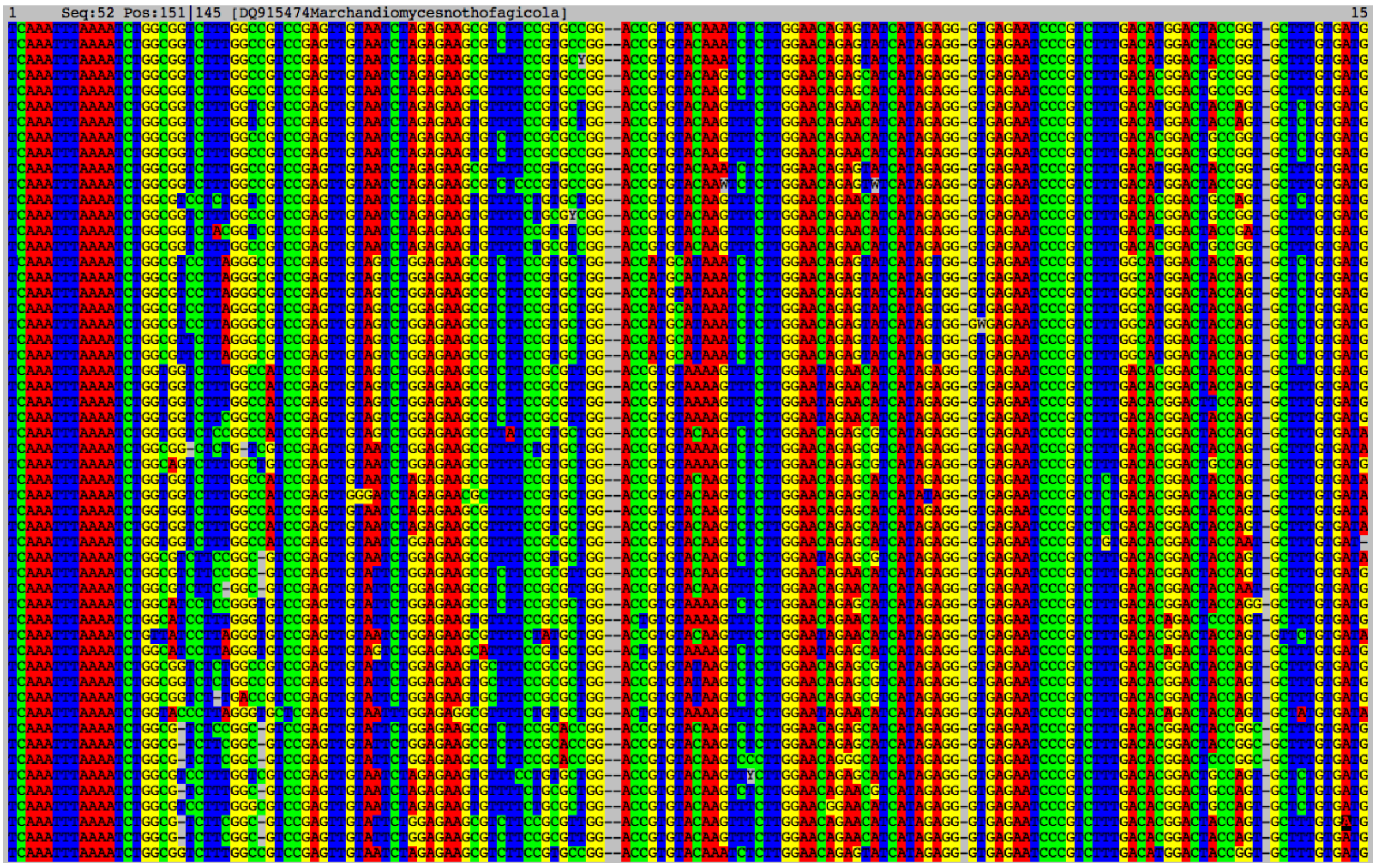

**Figure 3. a)** A satisfactory multiple sequence alignment run through MAFFT and then edited manually. The alignment is a subset of the 55-taxon alignment of Ghobad-Nejhad et al. (2010).

b.

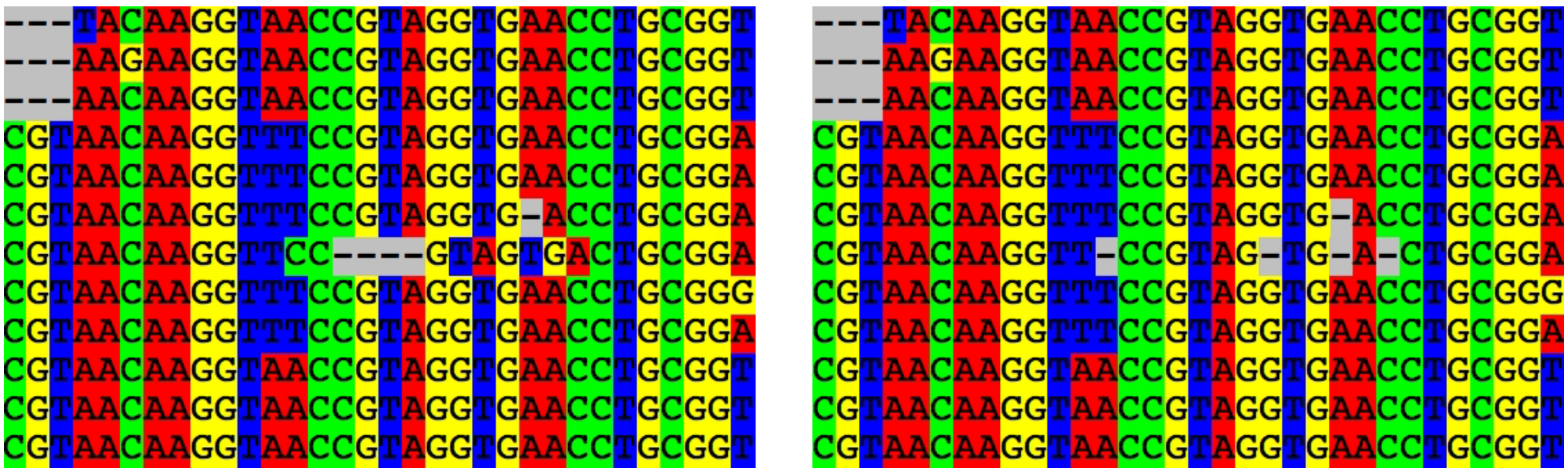

**Figure 3b)** Manual editing of a multiple sequence alignment in SeaView. The original alignment is given to the left, with the modified version to the right.

c.

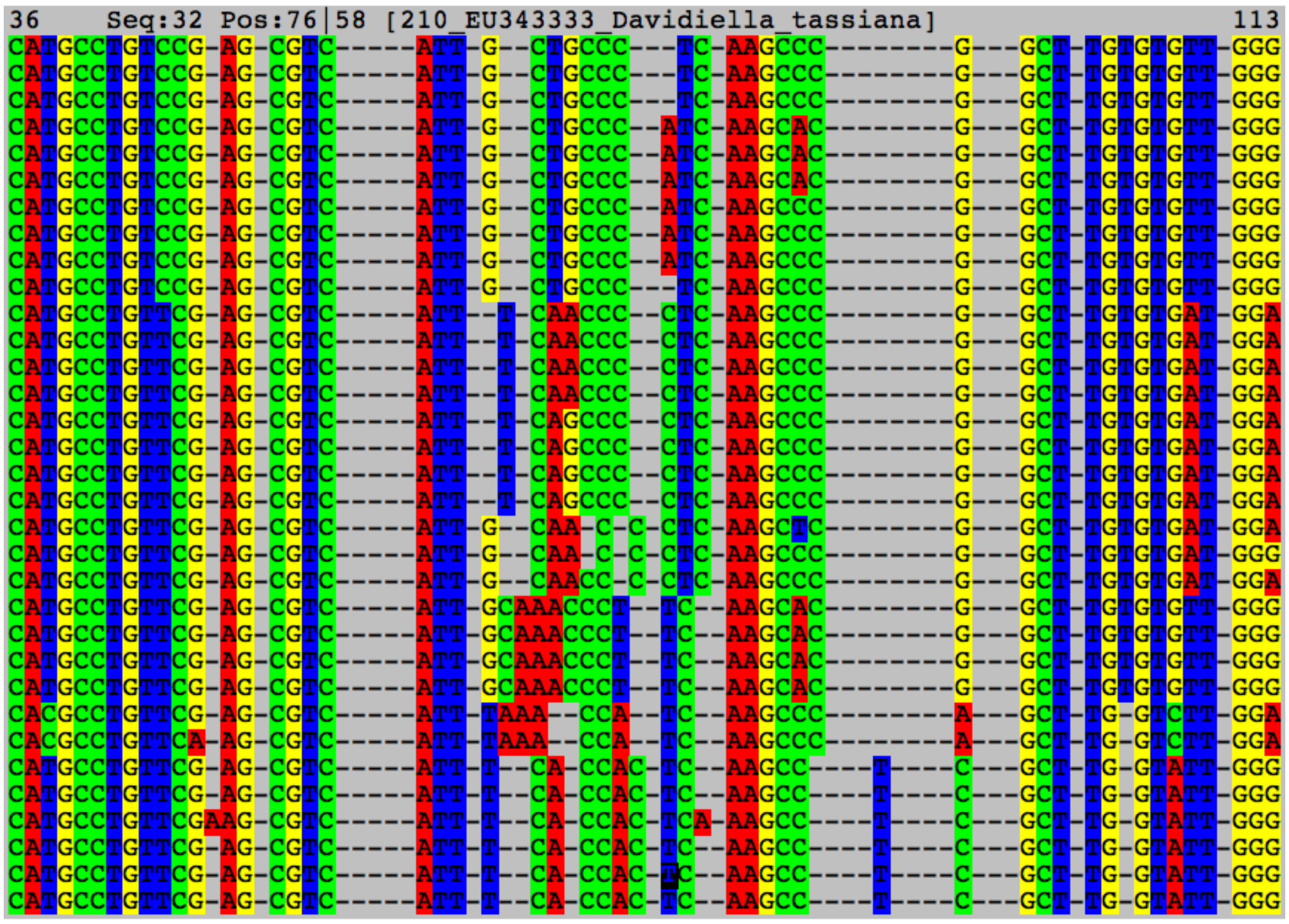

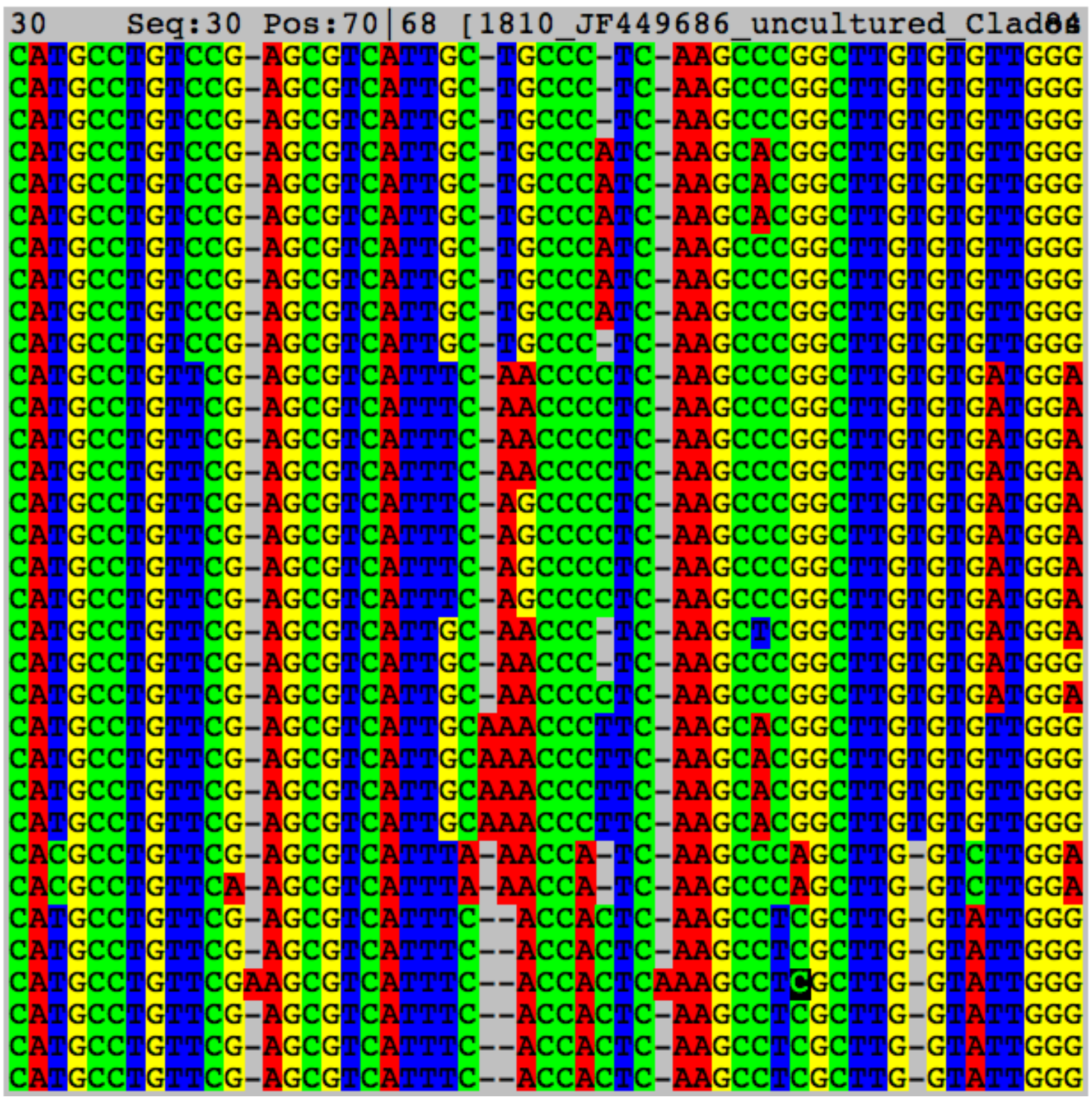

**Figure 3c)** Manual editing of a multiple sequence alignment in SeaView. The original alignment is given on top, with the modified version at the bottom with minimal gaps and ambiguities.

d.

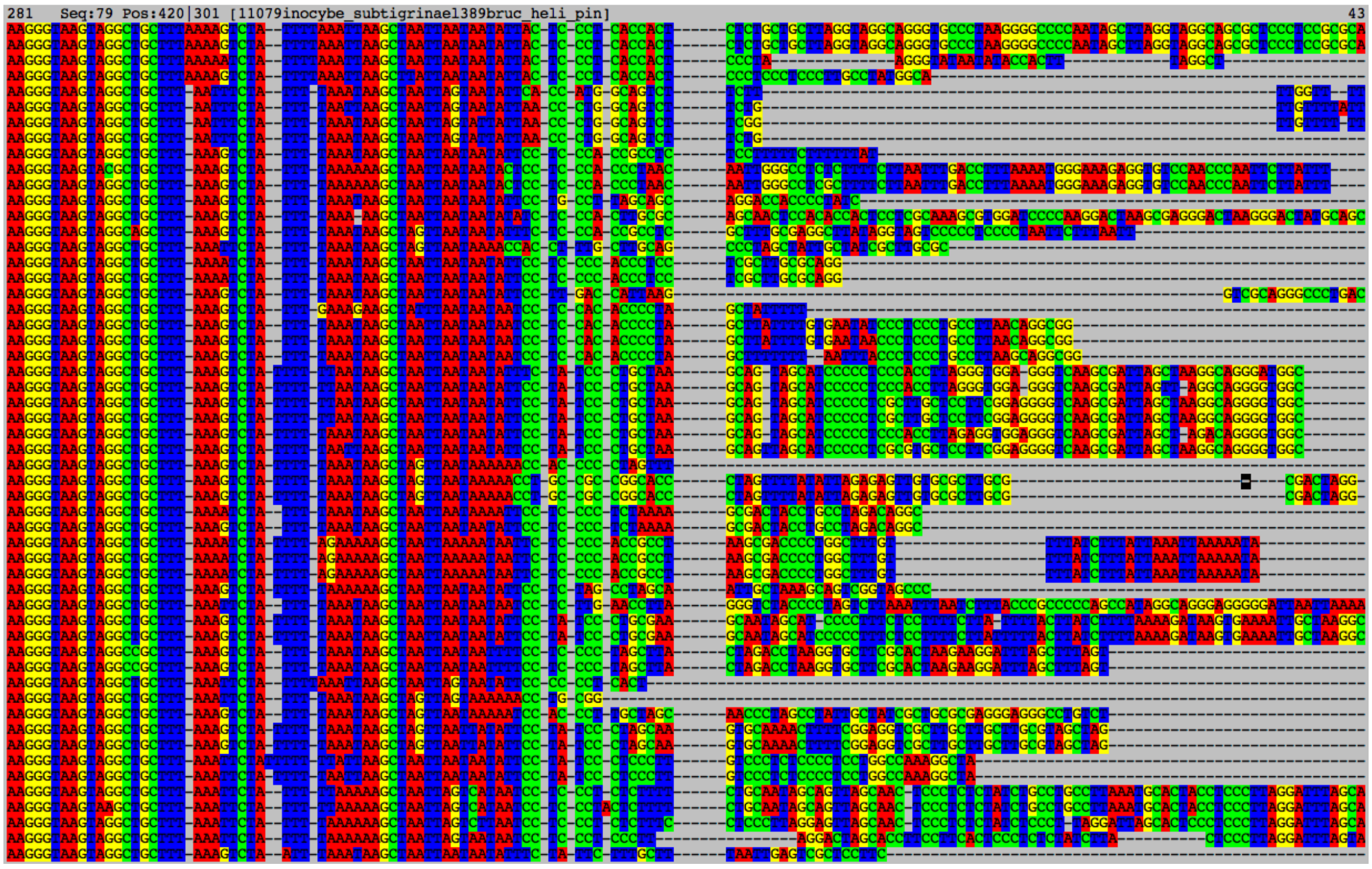

**d)** A portion of a multiple sequence alignment that defies scientifically meaningful global alignment. The first half of the alignment looks satisfactory, but the quality rapidly deteriorates midway through the alignment. Trying to shoehorn these sequence data into some sort of joint aligned form is sure to produce artificial, non-biological results. Such highly variable regions should be kept in the alignment for reference but excluded from the subsequent phylogenetic analysis. If and when the sequences are exported from the alignment, the user should make sure to employ the “Include all characters” option before exporting to avoid exporting only the parts used in the phylogenetic analysis. While the right half of the alignment is not fit for joint alignment covering all of the query sequences, it is clearly not without signal at a lower level, i.e., for subsets of the sequences.

*Phylogenetic analysis.* There are many different approaches to phylogenetic analysis, ranging from distance-based through parsimony to maximum likelihood and Bayesian inference (cf. Hillis et al. 1996, Felsenstein 2004, and Yang and Rannala 2012). For highly coherent, low-homoplasy datasets, most approaches are likely to produce similar results. That said, there is no single, optimal choice of analysis approach for all datasets, and it is a good idea for the user to start exploring their data using different methods. A fast, up-to-date program for parsimony analysis is TNT (Goloboff et al. 2008). If the users were to browse through various recent phylogeny-oriented publications in high-profile journals, they would probably come to the conclusion that Bayesian inference and (particularly for large datasets) maximum likelihood, both of which are explicitly model-based (parametric), have a lot of momentum at present. For one thing, it is widely accepted that they are better able to correct for spurious effects caused by long-branch attraction, which is a problem of major concern in phylogenetic inference (Bergsten 2005). Indeed, some journals are likely to require an analysis using one of these two methods. There are many different programs for phylogenetic analysis; a good list is found at http://evolution.genetics.washington.edu/phylip/software.html. Among the more popular tools for Bayesian analysis are MrBayes (Ronquist et al. 2012) and BEAST (Drummond et al. 2012); for maximum likelihood, RAxML (Stamatakis 2006) and GARLI (Zwickl 2006) see heavy use. Many of these phylogenetic analysis programs are furthermore available as web services on bioinformatics portals such as CIPRES (http://www.phylo.org/portal2/) and BioPortal (http://www.bioportal.uio.no/), which overcomes problems with limited computational capacity of most personal computers. When using explicitly model-based approaches, it is important to consider how the molecular evolution should be modelled, i.e., how the changes in the current sequence data have evolved. The above model-based programs allow, to various degrees, for implementation of separate models for different partitions (typically: each gene/marker) of the dataset. For example, in the case of the ITS region, the user should be aware that the two spacer regions (ITS1 and ITS2) and the intercalary 5.8S gene usually differ dramatically in substitution rate, so it may be appropriate to use different models of nucleotide evolution for each of these (as assessed through, e.g., jModelTest (Posada 2008)). It is also possible to test if a region is best modelled as one or separate partitions using PartitionFinder (Lanfear et al. 2012). Gaps are normally treated as missing data by phylogenetic inference programs; if the user feels that some gap-related characters are particularly phylogenetically informative (and particularly if the sequences under scrutiny are very similar such that informative characters are thinly seeded), the gap-related information can be added as a recoded, separate partition (see for example Simmons and Ochoterena 2000). Some approaches try to reconstruct the insertion/deletion and substitution processes simultaneously (Suchard and Redelings 2006; Varón et al. 2010).

As datasets grow beyond some 20 sequences, there are no analytical solutions for finding (guaranteeing) the best phylogenetic tree for any optimality criterion. Instead, heuristic searches - that is, searches that do not guarantee that the optimal solution is found - are employed to infer the tree or a distribution of trees. The most popular tools for Bayesian inference of phylogeny rely on Markov Chain Monte Carlo (MCMC) sampling, a heuristic sampling procedure where one to many searches (“chains”) traverse the space of possible trees (typically fully parameterized with, e.g., branch lengths) one step at the time, comparing the next tree (step ahead) only to the tree it presently holds in memory (Ronquist and Huelsenbeck 2003). There are several statistics that can be calculated to test if the chain has reached a steady state, i.e., if the chain has converged to a set of stable results for which significant improvements do not seem possible (Rambaut and Drummond 2007; Nylander et al. 2008). The set of trees (and parameters) inferred is then used to compute (e.g.) a majority-rule consensus tree with branch lengths and support values. At a more general level, there are methods to evaluate the reliability of tree topology and individual branches for all major approaches to phylogenetic inference. Branch support should routinely be estimated – regardless of approach – and reported on in the subsequent publication. The most common measures of branch support are Bayesian posterior probabilities (BPP; applicable in Bayesian inference) and, in distance/parsimony/likelihood-based inferences, non-parametric resampling methods such as traditional bootstrap (Felsenstein 1985) and jackknife (Farris et al. 1996). Bayesian posterior probabilities are estimated from the proportion of trees exhibiting the clade in question from the posterior distributions generated by the MCMC simulation and represent the probability that the corresponding clades are true conditional on the model and the data. Bootstrap/jackknife values are obtained from iterated resampling/dropping of characters from the multiple sequence alignment and re-running the phylogenetic analysis for each new alignment; the bootstrap/jackknife values then represent the proportion of times the corresponding clades were recovered from these perturbed alignments.

We argue that branch lengths should always be indicated in phylogenetic trees. In the context of parsimony, branch lengths represent the minimum number of mutations (steps) separating two nodes, whereas in maximum likelihood/Bayesian inference branch lengths are normally given in the unit of expected changes per site. Any extreme branch lengths observed for any of the taxa should be explored for mistakes in the alignment, sequences of poor quality, inclusion of taxa that are not closely related to the other taxa, and increased evolutionary rate along that branch. Such long-branched taxa call for a re-evaluation of the alignment and possibly also the taxon sampling; long branches represent one of the most difficult problems in phylogenetic inference (Bergsten 2005). If the user combines two or more genes from the same linkage group (e.g., the mitochondrial genome) in the alignment, it is customary to test the dataset for conflicts prior to undertaking the phylogenetic analysis (Hipp et al. 2004).

Interpretation of phylogenetic results and trees can be surprisingly tricky (Hillis et al. 1996; Felsenstein 2004). The first thing the user should check is that the sequence used as an outgroup really is an appropriate outgroup with respect to the ingroup sequences. The answer tends to come naturally when several progressively more distant sequences with respect to the ingroup are included in the alignment. Sequences of low read quality or of a chimeric nature tend to be found on unusually long branches or as isolated sister taxa to larger clades, and a second look at such sequences is always warranted (cf. Berney et al. 2004). Branches that do not receive significant, but rather modest, support can usually be thought of as non-existent such that what they really depict is the state of "no resolution available"; to draw far-reaching conclusions for such modestly supported clades is wishful thinking, that is, something the user should stay clear of. What constitutes "significant" branch support is a non-trivial question, though, and the user is advised to focus on clades that appear strongly supported (e.g., more than 90% bootstrap/jackknife or more than 0.95 BPP). In the context of clades and branches, it is tempting to identify some taxa as "basal" to others, but nearly all phylogenetic uses of the word "basal" are conceptually flawed (Krell and Cranston 2004), and the user is best off avoiding it altogether. The "sister clade/taxon" construct is the most straightforward alternative.

*Presenting phylogenetic results.* The end product of a phylogenetic analysis is typically a tree in the (text-based) Newick (http://en.wikipedia.org/wiki/Newick_format) or Nexus (Maddison et al. 1997) formats. This file can be loaded into tree viewing programs such as FigTree (http://tree.bio.ed.ac.uk/software/figtree/), manipulated, and saved in a graphics format (preferably a vector-based format such as .svg, .emf, or .eps). This file, in turn, can be loaded into, e.g., Corel Draw or Adobe Illustrator (GIMP and InkScape are free alternatives) for further processing, e.g., cleaning up the taxon names and highlighting focal clades (Figure 4). There is nothing wrong with presenting a phylogenetic tree in a straightforward, non-embellished style, particularly not for trees with a limited number of sequences. Many researchers nevertheless prefer to take their trees to the next level by, e.g., mapping morphological characters onto clades, indicating generic boundaries with colours, and collapsing large clades into symbolic units. iTOL (Letunic and Bork 2011), Mesquite (Maddison and Maddison 2011), and OneZoom (Rosindell and Harmon 2012) are powerful tools for such purposes. The Deep Hypha issue of Mycologia (December 2006) or the iTOL site (http://itol.embl.de/) may serve as sources of inspiration on how trees could be manipulated and processed to facilitate interpretation and highlight take-home messages.

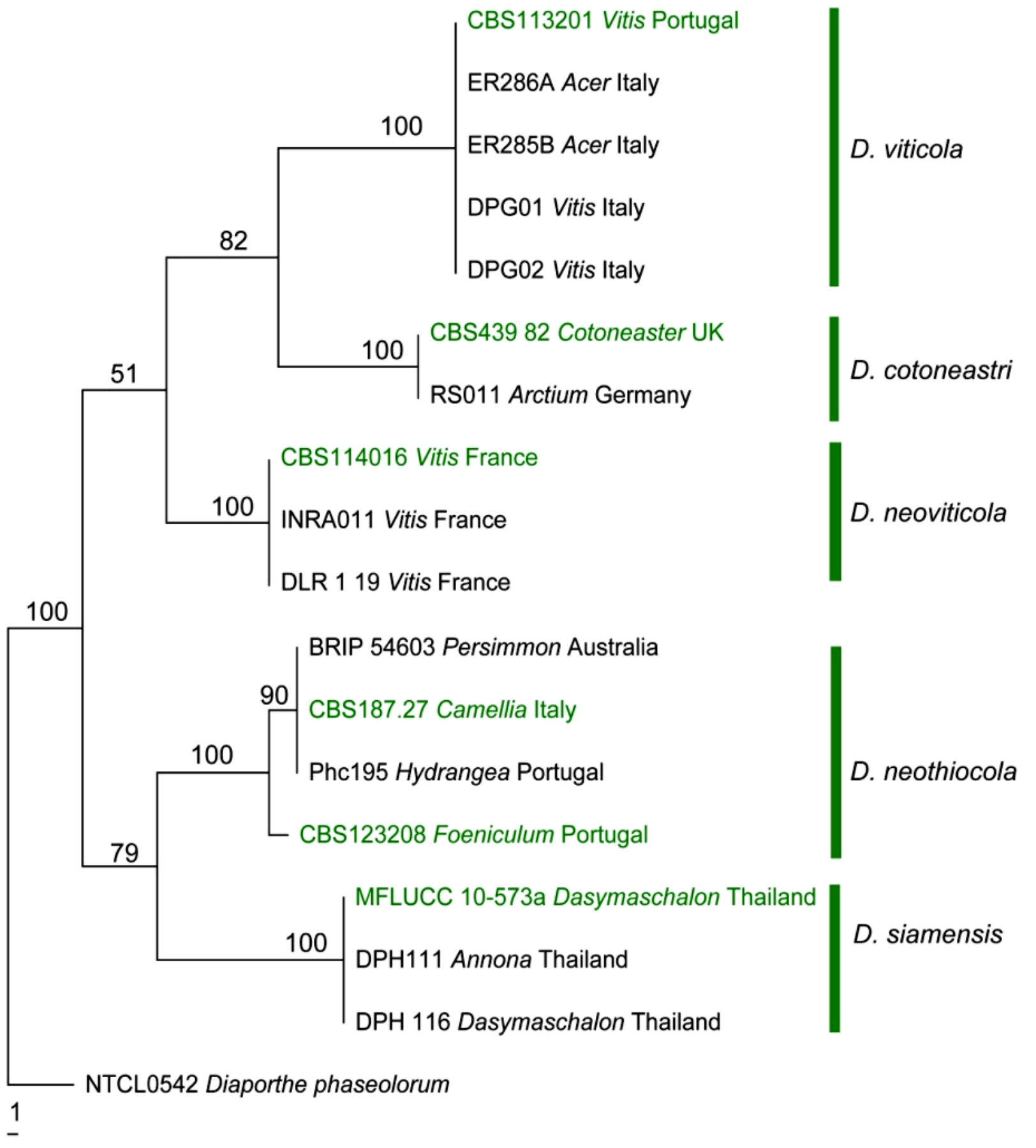

**Figure 4.** An example of identification of species of plant pathogenic fungi of the genus *Diaporthe* (*Phomopsis*) based on ex-type ITS sequences (Udayanga et al. 2012). Phylogram inferred from the parsimony analysis. One of the most parsimonious trees generated based on ex-type sequences and some unidentified strains for quick identification of species. Ex-type sequences are bold and green, and a random selection of isolates from various studies is included with original strain codes. Bootstrap support values exceeding 50% are shown above the branches. The tree is rooted with *Diaporthe phaseolorum*.

Most studies employing phylogenetic analysis are heavily underspecified in the Materials & Methods section and, worse, do not provide neither the multiple sequence alignment nor the phylogenetic trees derived as files (cf. Leebens-Mack et al. 2006). In our opinion, a good study specifies details on the multiple sequence alignment (e.g., number of sites in total, constant sites, and parsimony informative sites); on all relevant/non-default settings of the phylogeny program employed; and on the phylogenetic trees produced. All software packages should be cited with version number. The user should always bundle the multiple sequence alignment and the tree(s) with the article through TreeBase (http://www.treebase.org ; Sanderson et al. 1994), DRYAD (http://datadryad.org/ ; Greenberg 2009), or even as online supplementary items to the article (as applicable). This makes subsequent data access easy for the scientific community (cf. Mesirov 2010) and helps dispel the old assertion of taxonomy as a secretive, esoteric discipline.

## 5. The publication process

The scientific publication is something of the common unit of qualification in the natural sciences and the end product of many scientific projects. Having spent considerable time with the data collection and analysis, the user may be tempted to rush through the writing phase just to get the paper out. This is usually a mistake, because publishing tends to be more difficult, and to require more of the authors, than one perhaps would think. Indeed, it is not uncommon for a project to take two or more years from conception to its final, published state. Even very experienced researchers struggle with the writing phase, and we advise the user to start writing as early as possible. Three good ways to increase the chances of having the manuscript accepted in the end is to have at least two (external if possible) colleagues read through it well ahead of submission; to make sure that the language used is impeccable; and to follow the instructions of the target journal down to the very pixel. Many journals are flooded with submissions and are only too happy to reject manuscripts if they deviate ever so slightly from the formally correct configuration. A few general considerations follow below.

*Choice of target journal.* The user should decide upon the primary target journal before writing the first word of the manuscript. The second- and third-choice journals should ideally be chosen to be close in scope and style with respect to the primary one, so that the user would not have to spend significant time restructuring or refocusing the manuscript if the primary journal rejects it. The scope of the journal dictates how the manuscript should be written: if it is a more general (even non-mycological) journal, the user should probably focus on the more general, widely relevant aspects of the results. General journals have the advantage of reaching a broader audience than taxonomy-oriented journals, and if the user feels she has the data to potentially merit such a choice then she should certainly try. However, trying to shoehorn smaller taxonomic papers into more general journals is likely to prove a futile exercise. There is nothing wrong with publishing in more restricted journals, and to be able to tell the difference between a manuscript with the potential for a more general journal and a manuscript that probably should be sent to a more restricted journal is a skill that is likely to save the user considerable time and energy. The user should also be prepared to be rejected: this is a part of being in science and not something that should be taken too personally. That said, the user should scrutinize the rejection letter for clues to how the manuscript could be improved. The user should make it a habit to try to implement at least the easiest, and preferably several more, of those suggested changes in the manuscript before submitting it to the next journal in line.

Many funding agencies and institutional rankings ascribe extra weight to publications in journals that are indexed in the ISI Thompson Web of Science (http://thomsonreuters.com/), i.e., publications in journals that have, or are about to get, a formal impact factor (http://en.wikipedia.org/wiki/Impact_factor). Citations are often quantified in an analogous manner. This makes it a good idea to seek to target journals with a formal impact factor (or at least journals with an explicit ambition to obtain one), more or less irrespectively of what that impact factor is. We are under the impression that impact factors are falling out of favour as a bibliometric measurement unit of scientific quality – which would perhaps reduce the incentive for seeking to maximize the impact factor in all situations – but if the user has a choice, it may make sense to strive for journals that have an impact factor of 1 or higher. If given the choice between a journal run by a society and a journal run by a commercial publishing company – with otherwise approximately equal scope and impact – we would go for the one maintained by the society. A recent overview of journals with a full or partial focus on mycology is provided by Hyde and KoKo (2011).

*Open access.* An increasing number of funding agencies require that projects that receive funding publish all their papers (or otherwise make them available) through an open access model, i.e., freely downloadable (cf. http://www.doaj.org/). As a consequence, the number of open access journals has exploded during the last few years. Similarly, most non-open access journals with subscription fees now offer an "open choice" alternative where individual articles are made open access online. In both cases, there is normally a fee involved, and the fee tends to be sizable (e.g., US$1350 for PLoS ONE, US$1990 for the BMC series, and US$3000 for many Elsevier articles as of October 2012). Less well known is perhaps that most major publishing companies allow the authors to make pre- or post-prints (the first and the last version of the manuscript submitted to the journal, i.e., "pre" and "post" review)

available as a Word or PDF file on their personal homepages or certain public repositories such as arXiv (http://arxiv.org/). The SHERPA/RoMEO database at http://www.sherpa.ac.uk/romeo/ has the full details on what the major publishing companies/journals allow the authors to do with their pre- and post-print manuscripts. To make a manuscript publicly available in pre- or postprint form, in turn, qualifies as "open access" as far as many funding agencies are concerned. Thus, if the user needs to publish open access but cannot afford it, this may be the way to go.

There are some data to suggest that open access papers are cited more often than non-open access ones (MacCallum and Parthasarathy 2006), although the first integrative comparison based on mycological papers has yet to be undertaken. Most scientists, we imagine, are attracted to the ideas of openness, distributed web archiving, and of the dissemination of their results also to those who do not have access to subscription-based journals. Not all is gold that glitters, however. An increasing number of open access publishers may not have the authors' best interest in mind but are rather run as strictly commercial enterprises, often without direct participation of scientists. A Google search on "grey zone open access publishers" will produce lists of journals (and publishers) that the user may want to stay clear of. The papers are open access, but the peer review procedure tends to be less than stringent, and the journals seem to take little, if any, action to promote the results of the authors. Many of the journals are very poorly covered in literature databases and are unlikely to ever qualify for formal impact factors. It would seem probable that such publications would detract from, rather than add to, ones CV.

## 6. Other observations

Taxonomy is sometimes referred to as a discipline in crisis (Agnarsson and Kuntner 2007; Drew 2011). There is some truth to such claims, because the number of active taxonomists is in constant decline. Taxonomy furthermore struggles to obtain funding in competition with disciplines deemed more cutting-edge. However, genome-scale sequence data is already accessible for non-model organisms at a modest cost, and it can be anticipated that the standard procedures to obtain the molecular data described in this paper will in part be complemented and even replaced by NGS techniques at similar costs or less at some point in the not-too-distant future. This will require substantial bioinformatics efforts and data storage capabilities, but it also opens enormous possibilities for new scientific discoveries as well as more advanced and formalized models to trace phylogenies and the speciation process. Here we discuss some of the challenges faced by taxonomy in light of the project pursued by our imaginary user.

*Sanger versus next-generation sequencing.* The last seven years have seen dramatic improvements of, and additions to, the assortment of DNA sequencing technologies. Collectively referred to as next-generation sequencing (NGS), these new methods can produce millions – even billions – of sequences in a few days. At the time of writing this article, most NGS technologies on the market produce sequences of shorter length than those obtained through traditional Sanger sequencing, but this too is likely to change in the near future. The user should keep in mind, though, that the NGS techniques and Sanger sequencing are used for different purposes. NGS methods are primarily used to sequence genomes/RNA transcripts and to explore environmental substrates such as soil and the human gut for diversity and functional processes. To date there is no NGS technique to fully replace Sanger sequencing for regular research questions in systematics and taxonomy (neither in terms of read quality nor focus on single specimens), so it cannot be claimed that the present user employs "obsolete" sequencing technology. On the contrary, we feel the user should be congratulated for doing phylogenetic taxonomy, for even the most cutting-edge NGS efforts require names and a classification system from systematic and taxonomic studies in the form of, e.g., lists of species and higher groups recovered in their samples. Underlying those lists are reliable reference sequences and phylogenies produced by taxonomic researchers.

There is nevertheless a disconnect between NGS-powered environmental sequencing and taxonomy. Most environmental sequencing efforts recover a substantial number of operational taxonomic units (Blaxter et al. 2005) that cannot be identified to the level of species, genus, or even order (cf. Hibbett et al. 2011). However, since NGS sequences are not readily archived in INSD (but rather stored in a form not open to direct query in the European Nucleotide Archive; Amid et al. 2012), these new (or at least previously unsequenced) lineages are not available for BLAST searches and thus generally ignored. Similarly there is no device or centralized resource to database the fungal communities recovered in the many NGS-powered published studies, and most opportunities to compare and correlate taxa and communities across time and space are simply missed. Partial software solutions are being worked upon in the UNITE database (Abarenkov et al. 2010, 2011) and elsewhere, but we have no solution at present. It would be beneficial if all studies of eukaryote/fungal communities were to involve at least one fungal systematician/taxonomist. Conversely, fungal systematicians/taxonomists should seek training in bioinformatics and biodiversity informatics to become prepared to deal with the research questions involving taxonomy, systematics, ecology, and evolution in an increasingly connected, molecular world.

*How can we do systematics and taxonomy in a better way?* An important part of environmental sequencing studies is to give accurate names to the sequences. By recollecting species, designating epitypes, depositing type sequences in GenBank, and publishing phylogenetic studies/classifications of fungi, taxonomists provide a valuable service for present and future scientific studies. For example, it was previously impossible to accurately name common endophytic isolates of *Colletotrichum*, *Phyllosticta*, *Diaporthe* or *Pestalotiopsis* through BLAST searches, as the chances were very high that the names tagged to sequences in INSD for these genera were highly problematic (see KoKo et al. 2011). However, now that inclusive phylogenetic trees have been published for these genera (Glienke et al. 2011, Cannon et al. 2012, Udayanga et al. 2012, Maharachchikumbura et al. 2012) using types and epitypes, it is possible to name some of the endophytic isolates in these genera with reasonable accuracy. Some of the modern monographs with molecular data and compilations of different genera (e.g., Lombard et al. 2010, Bensch et al. 2010, Liu et al. 2012, and Hirooka et al. 2012) have become excellent guides for aspiring researchers with the knowledge on multiple taxonomic disciplines. The relevance of taxonomy will therefore be upgraded as long as care is taken to provide the sequence databases with accurately named, and richly annotated, sequences.

Taxonomists should also include aspects of, e.g., ecology and geography in taxonomic papers and seek to publish them where they are seen by others. We feel that a good taxonomic study should draw from molecular data (as applicable) and that it should seek to relate the newly generated sequence data to the data already available in the public sequence databases and in the literature. When interesting patterns and connections are found, the modern taxonomist should pursue them, including writing to the authors of those sequences to see if there is in fact an extra dimension to their data. As fellow taxonomists, we should all strive to help each other in maximizing the output of any given dataset. Taxonomists have a reputation of working alone - or at best in small, closed groups - and judging by the last few years' worth of taxonomic publications, that reputation is at least partially true. It would do taxonomy well if this practise were to stop. We envisage even arch-taxonomical papers such as descriptions of new species as opportunities to involve and integrate, e.g., ecologists and/or bioinformaticians into taxonomic projects. There are only benefits to inviting (motivated) researchers to ones studies: their focal aspects (such as ecology) will be better handled in the manuscript compared to if the author had handled them on their own, and the new co-authors are furthermore likely to improve the general quality of the manuscript by bringing an outside perspective and experience. We feel that perhaps 1-2 days of work is sufficient to warrant co-authorship.

On a related note, taxonomists should take the time to revisit their sequences in INSD to make sure their annotations and metadata are up-to-date and as complete as possible. At present this does not seem to be the case; Tedersoo et al. (2011) showed, for instance, that a modest 43% of the public fungal ITS sequences were annotated with the country of origin. We believe most researchers would agree that the purpose of sequence databases is to be a meaningful resource to science. The databases will however never be a truly meaningful resource to science if all data contributors keep postponing the updates of their entries to a day when free time to do those updates, magically, becomes available. Keeping ones public sequences up-to-date must be made a prioritized undertaking; taxonomy should be a discipline that shares its progress with the rest of the world in a timely fashion. If and when generating ITS sequences, fungal taxonomists should try to meet the formal barcoding criteria (http://barcoding.si.edu/pdf/dwg_data_standards-final.pdf; Schoch et al. 2012). This lends extra credibility to their work and is, in turn, likely to lead to a wider dissemination of their results. Taxonomists should similarly take the opportunity to add further value to their studies by including additional illustrations and other data in their publications. Space normally comes at a premium in scientific publications, but most journals would be glad to bundle highlights of additional interesting biological properties and additional figures of, e.g., fruiting bodies, habitats, and collection sites as supplementary, online-only items (cf. Seifert and Rossman 2010).

*Promoting ones results and career advancement.* Even if the authors' new scientific paper is excellent, the chances are that it will not stand out above the background noise to get the attention it deserves – at least not at a speed the author deems rapid enough. The author may therefore want to do some PR for their article. One step in that direction is to cite their paper whenever appropriate. Another step is to send polite, non-invasive emails to researchers that the author feels should know about the article – as long as the author provides a URL to the article rather than attaching a sizable PDF file, nobody is likely to take significant offence from such emails if sent in moderation. A third step would be to present the results at the next relevant conference. In fact, the authors could send a poster and article reprints to conferences they are not able to attend through helpful co-authors or colleagues. On a more general level, conferences form the perfect arena where to meet people and forge connections. We sincerely hope that all Ph.D. students in mycology will have the opportunity to attend at least 2-3 conferences (international ones if at all possible). Poster sessions and "Ph.D. mixers" are particularly relaxed events where friends and future collaborators can be established. Even highly distinguished, prominent mycologists tend to be open to questions and discussion, and many of them take particular care to talk to Ph.D. students and other emerging talents. We acknowledge, however, that not everyone is in a position to go to conferences. Fortunately we live in a digital world where friendly, polite emails can get you a long way. Indeed, several of the present authors have made their most valuable scientific connections through email only. Similarly, to sign up in, e.g., Google Scholar to receive email alerts when ones papers are cited is a good way to scout for researchers with similar research interests. We believe that the user should make an effort to connect to others, because it is though others that opportunities present themselves. Furthermore, if one chooses those "others" with some care, the ride will be a whole lot more enjoyable.

**Concluding remarks**

The present publication seeks to lower the learning threshold for using molecular methods in fungal systematics. Each header deserves its own review paper, but we have instead provided an topic overview and the references needed to commence molecular studies. We do not wish to oversimplify molecular mycology; at the same time, we do not wish to depict it as an overly complex undertaking, because in most cases it is not. Our take-home message is that the best way to learn molecular methods and know-how is through someone who already knows them well. To attempt to get everything working on one's own, using nothing but the literature as guidance, is a recipe for failure. To find people to commit to helping you may be difficult, however, and we advocate that anyone who contributes 1-2 days or more of their time towards your goals should be invited to collaborate in the study. Newcomers to the molecular field have the opportunity to learn from the experience – and past mistakes – of others. In this way they will get most things correct right from the start, and in this paper we have highlighted what we feel are best practises in the field. We believe that taxonomists should adhere to best practises and to maximise the scientific potential and general usefulness of their data, because taxonomy is a discipline that is central to biology and that faces multiple far-reaching, but also promising, challenges. It is however impossible to give advice that would hold true in all situations and throughout time, and we hope that the reader will use our material as a realization rather than as a binding recipe.

**Acknowledgements**

RHN acknowledges financial support from Swedish Research Council of Environment, Agricultural Sciences, and Spatial Planning (FORMAS, 215-2011-498) and from the Carl Stenholm Foundation. Bengt Oxelman is supported by a grant (2012-3719) from the Swedish Research Council. Manfred Binder is acknowledged for valuable discussions. Support from the Gothenburg Bioinformatics Network is gratefully acknowledged.